# Astrodynamical Space Test of Relativity using Optical Devices I (ASTROD I) - A class-M fundamental physics mission proposal for Cosmic Vision 2015-2025

Thierry Appourchaux
*Institut d'Astrophysique Spatiale, Centre Universitaire d'Orsay, 91405 Orsay Cedex, France*

Raymond Burston
*Max-Planck-Institut für Sonnensystemforschung, 37191 Katlenburg-Lindau, Germany*

Yanbei Chen
*Department of Physics, California Institute of Technology, Pasadena, California 91125, USA*

Michael Cruise
*School of Physics and Astronomy, Birmingham University, Edgbaston, Birmingham, B15 2TT, UK*

Hansjörg Dittus
*Centre of Applied Space Technology and Microgravity (ZARM), University of Bremen, Am Fallturm, 28359 Bremen, Germany*
e-mail: dittus@zarm.uni-bremen.de

Bernard Foulon
*Office National D'Édudes et de Recherches Aerospatiales, BP 72 F-92322 Chatillon Cedex, France*

Patrick Gill
*National Physical Laboratory, Teddington TW11 0LW, United Kingdom*

Laurent Gizon
*Max-Planck-Institut für Sonnensystemforschung, 37191 Katlenburg-Lindau, Germany*

Hugh Klein
*National Physical Laboratory, Teddington TW11 0LW, United Kingdom*

Sergei Klioner
*Lohrmann-Observatorium, Institut für Planetare Geodäsie, Technische Universität Dresden, 01062 Dresden, Germany*

Sergei Kopeikin
*Department of Physics and Astronomy, University of Missouri, Columbia, Missouri 65221, USA*

Hans Krüger, Claus Lämmerzahl
*Centre of Applied Space Technology and Microgravity (ZARM), University of Bremen, Am Fallturm, 28359 Bremen, Germany*

Alberto Lobo
*Institut d´Estudis Espacials de Catalunya (IEEC), Gran Capità 2-4, 08034 Barcelona, Spain*

Xinlian Luo
*Department of Astronomy, Nanjing University, Nanjing, 210093, China*

Helen Margolis
*National Physical Laboratory, Teddington TW11 0LW, United Kingdom*

Wei-Tou Ni, Antonio Pulido Patón
*Center for Gravitation and Cosmology, Purple Mountain Observatory, Chinese Academy of Sciences, No. 2, Beijing W. Road, Nanjing, 210008, China*
e-mail: wtni@pmo.ac.cn

Qiuhe Peng
*Department of Astronomy, Nanjing University, Nanjing, 210093, China*

Achim Peters
*Department of Physics, Humboldt-University Berlin, 10117 Berlin, Germany*

Ernst Rasel
*Leibniz Universitaet Hannover, Institut für Quantenoptik, Welfengarten 1, 30167 Hannover, Germany*

Albrecht Rüdiger
*Max-Planck-Insitut für Gravitationsphysik, 30167 Hannover, Germany*

Étienne Samain
*Observatoire de la Côte d'Azur, UMR Gemini, R&D Métrologie, 06460 Caussols, France*

Hanns Selig
*Centre of Applied Space Technology and Microgravity (ZARM), University of Bremen, Am Fallturm, 28359 Bremen, Germany*

Diana Shaul, Timothy Sumner
*High Energy Physics Group, Blackett Laboratory, Imperial College London, Prince Consort Road, London, SW7 2BZ, UK*

Stephan Theil
*Centre of Applied Space Technology and Microgravity (ZARM), University of Bremen, Am Fallturm, 28359 Bremen, Germany*

Dr. Pierre Touboul
*Office National D'Édudes et de Recherches Aerospatiales, BP 72 F-92322 Chatillon Cedex, France*

Dr. Slava Turyshev
*Jet Propulsion Laboratory, California Institute of Technology, Pasadena, California 91109, USA*

Haitao Wang
*Nanjing University of Aeronautics and Astronautics, Nanjing, 210016, China*

Li Wang
*Deep Space Exploration & Space Science Technology Research Division, Research & Development Center, China Academy of Space Technology, Beijing, 100094, China*

Linqing Wen
*School of Physics,University of Western Australia, Crawley, WA 6009, Australia*

Andreas Wicht
*Institut für Experimentalphysik, Heinrich-Heine-Universität, 40225Düsseldorf, Germany*

Ji Wu
*Center for Space Science and Application Research, Chinese Academy of Sciences, Beijing, China*

Xiaomin Zhang
*DFH Satellite Co. Ltd., No. 104, Youyi Rd., Haiden District, Beijing, 100094, China*

Cheng Zhao
*Center for Gravitation and Cosmology, Purple Mountain Observatory, Chinese Academy of Sciences, No. 2, Beijing W. Road, Nanjing, 210008, China*

**Abstract:** ASTROD I is a planned interplanetary space mission with multiple goals. The primary aims are: to test General Relativity with an improvement in sensitivity of over 3 orders of magnitude, improving our understanding of gravity and aiding the development of a new quantum gravity theory; to measure key solar system parameters with increased accuracy, advancing solar physics and our knowledge of the solar system; and to measure the time rate of change of the gravitational constant with an order of magnitude improvement and the anomalous Pioneer acceleration, thereby probing dark matter and dark energy gravitationally. It is an international project, with major contributions from Europe and China and is envisaged as the first in a series of ASTROD missions. ASTROD I will consist of one spacecraft carrying a telescope, four lasers, two event timers and a clock. Two-way, two-wavelength laser pulse ranging will be used between the spacecraft in a solar orbit and deep space laser stations on Earth, to achieve the ASTROD I goals.

A second mission, ASTROD (ASTROD II) is envisaged as a three-spacecraft mission which would test General Relativity to one part per billion, enable detection of solar g-modes, measure the solar Lense-Thirring effect to 10 parts per million, and probe gravitational waves at frequencies below the LISA bandwidth. In the third phase (ASTROD III or Super-ASTROD), larger orbits could be implemented to map the outer solar system and to probe primordial gravitational-waves at frequencies below the ASTROD II bandwidth.

**1. Introduction and mission summary**

The general concept of ASTROD (Astrodynamical Space Test of Relativity using Optical Devices) is to have a constellation of drag-free spacecraft (S/C) navigate through the solar system and range with one another and/or ground stations using optical devices to map and monitor the solar-system gravitational field, to measure related solar-system parameters, to test relativistic gravity, to observe solar g-mode oscillations, and to detect gravitational waves. The gravitational field in the solar system influences the ranges and is determined by three factors: the dynamic distribution of matter in the solar system; the dynamic distribution of matter outside the solar system (galactic, cosmological, etc.); and gravitational waves propagating through the solar system. Different relativistic theories of gravity make different predictions of the solar-system gravitational field. Hence, precise measurements of the solar-system gravitational field test these relativistic theories, in addition to gravitational wave observations, determination of the matter distribution in the solar-system and determination of the observable (testable) influence of our galaxy and cosmos. Since ranges are affected by gravitational fields, from the precise range observations and their fitting solution, the gravitational field in the solar system with different contributing factors can be determined with high precision. Two keys in this mission concept are range measurement accuracy and minimizing spurious non-geodesic accelerations. The range accuracy depends on timing uncertainty, while minimizing spurious non-geodesic accelerations relies on drag-free performance.

A baseline implementation of ASTROD (also called ASTROD II) is three-spacecraft mission with two spacecraft in separate solar orbits and one spacecraft near Earth (Sun-Earth L1/L2 point), each carrying a payload of a proof mass, two telescopes, two 1–2 W lasers, a clock and a drag-free system and range coherently with one another using lasers [1, 2]. ASTROD I using one spacecraft ranging with ground laser stations is the first step toward the baseline mission ASTROD [3, 4]. In our earlier version of ASTROD I mission concept, we used both laser pulse ranging and laser interferometric ranging and assumed a timing uncertainty of 10 ps together with residual acceleration noise requirement of

$$S_{\Delta a}^{1/2}(f) = 3.3 \times 10^{-14} [(0.3 \text{ mHz}/f + 30 \times (f/3 \text{ mHz})^2] \text{ m s}^{-2} \text{ Hz}^{-1/2}, \qquad (1)$$

over the frequency range of 0.1 mHz < $f$ < 100 mHz with $f$ in unit of mHz [3, 4]. The laser pulse ranging and laser interferometric ranging use two different wavelengths to serve as 2-color laser ranging to measure and subtract atmospheric refraction delay; the interferometric ranging would also be a technological preparation for the later interferometric ASTROD missions. In the Comic Vision ASTROD I proposal, for making it technologically simpler we trimmed the ranging method to laser pulse ranging (with two colors) only. As to the timing accuracy, we used the accuracy of already developed event timer, i.e., 3 ps. From the experience of GOCE (Ocean Circulation Explorer) sensor heads [5, 6], a 3.3-fold more stringent noise requirement in the inertial sensor / accelerometer compared with that of (1) was deemed feasible in the time frame of the proposal. The range measurement accuracy is 0.9 mm (3 ps); compared with the maximum S/C distance to Earth of 1.7 AU, the fractional accuracy is $0.3 \times 10^{-14}$ in range. This is more than 3 orders of magnitude improvement over radio range experiment which has an accuracy of a couple of meters at present. Test of relativistic gravity and measurement of solar system parameters will be improved by more than 3 orders of magnitude. The corresponding improvement in drag-free performance is required in company to achieve this improvement.

ASTROD I mission concept has one spacecraft carrying a payload of a telescope, four lasers (two spares), and a clock ranging with ground stations (ODSN: Optical Deep Space Network). A schematic of the payload configuration of ASTROD I is shown in Fig. 1. The cylindrical spacecraft with diameter 2.5 m and height 2 m has its surface covered with solar panels. In orbit, the cylindrical axis is perpendicular to the orbital plane with the telescope pointing toward the ground laser station. The effective area to receive sunlight is about 5 m$^2$ and this can generate over 500 W of power. The total mass of the spacecraft is 490 kg. That of the payload is 160 kg with a science data rate of 500 bps. The laser ranging is between a fiducial point in the spacecraft and a fiducial point in the ground laser station. The fiducial point in the spacecraft can be a separate entity which has a definite positional relation with respect to the proof mass. Incoming pulsed laser light will be collected using a 300 mm diameter f /1 Cassegrain telescope. This telescope will also transmit pulsed laser light from the spacecraft to Earth. For the ground laser stations, we can use the present lunar laser ranging (LLR) stations or large satellite laser ranging (SLR) stations. The time epochs of transmitting and receiving of laser pulses in the ground stations and in the spacecraft is recorded by event timers using precision clocks. An event timer has been designed at OCA (Observatoire de la Côte d'Azur) in the framework of both T2L2 (Time Transfer by Laser Link) project and the laser ranging activities [7, 8]. At present, the prototype of the OCA timer is fully operational having a precision less than 3 ps, a linearity error of 1 ps rms and a time stability less than 0.01 ps over 1000 s with dead time less than 10 μs. T2L2 on board Jason 2 was launched on June 20, 2008 which overlaps with the Jason-1 mission to secure the continuity of high accuracy satellite altimetry observations [9]. Preliminary analysis of data confirms anticipation. With a cesium clock on board spacecraft, a fractional precision of distance determination of $10^{-14}$ will be achieved.

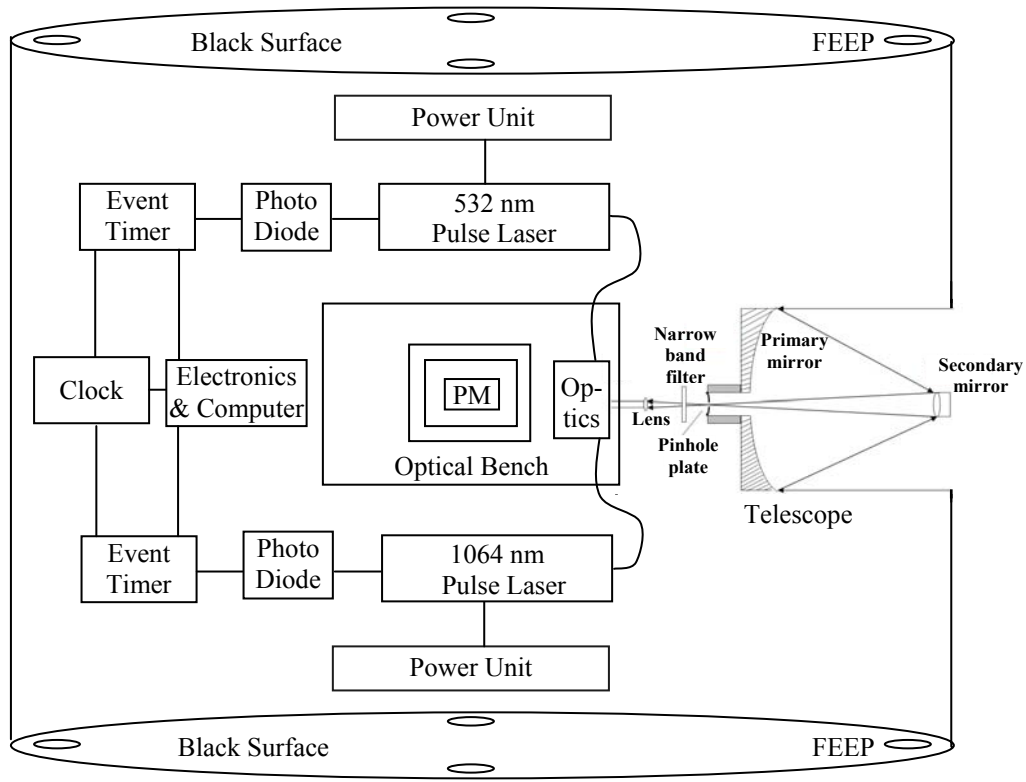

**Fig. 1** A schematic diagram of paylowad configuration for ASTROD I. (PM: Proof Mass)

The scientific aims include a better measurement of the relativistic parameters and many solar system parameters. For a launch in 2017, after two encounters with Venus, the orbital period can be shortened to 165 days. The science measurement starts immediately after orbit correction [3, 4]. The range measurement can be performed a couple of times a day or one time in a couple of days. Since the measurement principle does not depend on cancellation, some deviation from the fiducial orbit design is allowed. The two Venus flybys are for shortening the time to reach the other side of the Sun (and to measure the Venus gravity field). After about 370 days from launch, the spacecraft will arrive at the other side of the Sun for a precise determination of the light deflection parameter γ through Shapiro time delay measurement so that the correlation of this parameter and the relativistic parameter nonlinear parameter β can be decreased. The configuration after about 370 days from launch is sketched in Figure 2. A mission summary is compiled in Table 1.

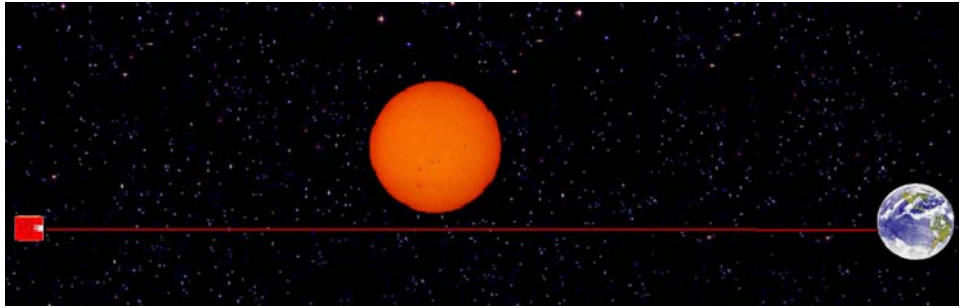

**Fig. 2** A sketch of ASTROD I mission at about 370 days after launch (Sun located near the line between ASTROD I and earth). The spacecraft is about 1.7 AU from Earth at this time.

**Table 1 ASTROD I Mission Summary**

| | |
|---|---|
| Objective | - Testing relativistic gravity and the fundamental laws of spacetime with more than three order-of-magnitude improvement in sensitivity<br>- Initiating a revolution of astrodynamics with laser ranging in the solar system<br>- Increasing the sensitivity of solar, planetary and asteroid parameter determination by 1 to 3 orders of magnitude. |
| Payload | - **Laser systems for pulse ranging** one 532 nm (plus 1 spare) and one 1064 (plus 1 spare) pulsed Nd:YAG laser with timing device for recording the transmitting time of space laser pulse and the receiving time of the incoming laser pulse from ground laser stations (range accuracy for a typical normal point: 3 ps (0.9 mm))<br>- **Avalanche photodiode array**<br>- **300 mm diameter f/1 Cassegrain telescope** (transimit/receive), λ/4 outgoing wavefront quality<br>- **Drag-free proof mass**<br>    o 50 x 50 x 35 mm$^3$ rectangular parallelepiped<br>    o Au-Pt alloy of extremely low magnetic susceptibility (< 5 x 10$^{-5}$)<br>    o Ti-housing at vacuum < 10$^{-5}$ Pa<br>    o six-degree-of-freedom capacitive sensing<br>- **Sunlight shield system**<br>- **Event Timer**<br>- **Cesium/Rubidium clock** |
| Ground laser Stations | 1.2 m (Yunnan) and 1.5 m (CERGA) diameter telescopes (transmit/receive) |
| Orbit | Launch via a large elliptic earth transfer orbit with perigee between 200 km and 300 km and apogee about 500,000 km to solar orbit with orbit period 300 days with a 100 kg propulsion unit to be separated after usage. The initial orbit is corrected using a medium-sized ion thruster. S/C is on drag-free and collects ranging data. After two encounters with Venus to get gravity-assistance the orbit period of the spacecraft SC can be decreased to 165 days. The apparent position of S/C reaches the opposite side of the Sun shortly after 360 days, 670 days and 1050 days from launch. |
| Launcher: | Long March IV B (CZ-4B) |
| Spacecraft: | 3-axis stabilized drag-free spacecraft |
| (total) mass: | 490 kg (S/C including a medium-size ion thruster for orbit injection correction and a 100 kg apogee motor including liquid fuel) |
| (total) power: | 390 W |
| Drag-free performance: | 3 x10$^{-14}$ ms$^{-2}$/√Hz around 100 µHz for 2-sensitive axes in the orbital plane and 3 x10$^{-13}$ ms$^{-2}$/√Hz around 100 µHz for the 3$^{rd}$ axis, normal to orbital plane |
| Pointing accuracy: | 2 µrad |
| Payload mass: | 160 kg |
| Payload power: | 280 W |
| Science data rate: | 500 bps |
| Telemetry: | 5 kbps, for about 9 hours in two days |
| Gound station: | Deep Space Stations |
| Mission lifetime | 3 years (nominal); 8 years (extended) |

For ASTROD I, the critical technologies are (i) drag-free control; (ii) interplanetary pulse laser ranging and (iii) sunlight shielding. The drag-free requirement for ASTROD I is relaxed by one order of magnitude as compared with that of LISA (Laser Interferometer Space Antenna for gravitational-wave detection) [10]. The drag-free technologies under development for LISA Pathfinder [11] will largely meet the requirement of ASTROD I. Millimetre precision has already been achieved in lunar laser ranging [12, 13]. Interplanetary pulse laser ranging (both up and down) was demonstrated by MESSENGER, using its laser altimeter in 2005 [14]. The technologies needed for a dedicated mission using interplanetary pulse laser ranging with millimetre accuracy are already mature. Sunlight shielding is a common technology which needs to be developed for optical missions measuring Shapiro time delay and light deflection due to the Sun. The technological readiness of ASTROD I is relatively high and it could fit into ESA's proposed Drag-Free Fundamental Physics Explorer Mission Series.

We present scientific objectives in section 2, mission profile in section 3 and proposed payload instruments in section 4. In section 5, section 6, and section 7, we present basic spacecraft key factors, science operation and archiving, and outlook respectively.

## 2. Scientific Objectives

*The scientific goals of ASTROD I are threefold.* The first goal is to test relativistic gravity and the fundamental laws of spacetime with an improvement of more than three orders of magnitude in sensitivity, specifically, to measure the PPN (Parametrized Post-Newtonian) parameter γ (Eddington light-deflection parameter; for general relativity, it is 1) via Shapiro time delay to $3 \times 10^{-8}$, β (relativistic nonlinear-gravity parameter; for general relativity, it is 1) to $3 \times 10^{-8}$ and others with significant improvement; to measure the fractional time rate of change of the gravitational constant (dG/dt)/G with one and half orders of magnitude improvement; and to measure deviations from the Einsteinian gravitational acceleration, i.e., an anomalous Pioneer acceleration $A_a$ [15], with several orders of magnitude improvement. The second goal is to initiate a revolution of astrodynamics with laser ranging in the solar system. The third goal is to increase the sensitivity of solar, planetary and asteroid parameter determination by 1 to 3 orders of magnitude. In this context, the measurement of solar quadrupole moment parameter $J_2$ will be improved by two orders of magnitude, i.e., to $10^{-9}$. Table 2 gives a summary of these objectives. The present accuracies of the parameters are listed in the second column of the table.

There are two basic ways in which the mysteries of dark energy and dark matter might be solved. One is the direct measurement and characterization of the energy and matter concerned. The second is to modify our theory of gravity. Let us reminisce about the discovery of the first observed effect of relativistic gravity by Le Verrier in 1859 [17]. By 1859, the orbit observation of Mercury reached $10^{-8}$ of Newtonian gravity and there were discrepancies between the observational data and theoretical predictions. Le Verrier attributed these discrepancies to an additional 38″ per century anomalous advance in the perihelion of Mercury 1. There were two schools of thought as to the actual reason for this discrepancy: (i) there was some unobserved matter - proposals included a planet Vulcan and intra-Mercurial matter; (ii) Newton's law of gravity needed modification. The solar quadrupole moment was also discussed as a possible cause. The successful solution awaited the development of general relativity. *At present, as we have the means to test relativistic gravity to $10^{-8}$ in the solar system, do we have the potential to contribute to the solution of the dark energy-dark matter problem?*

The PPN [Parametrized Post-Newtonian] parameter γ is presently measured to be γ = 1.000021 ± 0.000023 by the Cassini experiment [18]. However, multiple tensor-scalar gravity theories with attractor predict a lower bound for the present value of this parameter at the level of (1 - γ) ~ $10^{-6}$ - $10^{-7}$ [19-21]. Our projected accuracy of $3 \times 10^{-8}$ for ASTROD I will completely cover this range and will be able to give a clear experimental resolution. The measurement of PPN parameter β to $3 \times 10^{-8}$ also enables the discrimination between different theories of gravity. In the measurement to $3 \times 10^{-8}$, 2PN (post-post-Newtonian) framework together with corresponding ephemeris is needed for data fitting; work is in progress on this aspect. When analyses are done in the full PPN framework or the extended 2PN framework, the uncertainties in other relativistic-gravity parameters will also be improved significantly and 2PN parameters measured for the first time.

The ASTROD I mission is capable of detecting a gravitomagnetic field caused by the angular momentum of the Sun. The gravitomagnetic field, according to Einstein's general theory of relativity, arises from moving matter (mass current) just as the magnetic field arises in Maxwell's theory from moving electric charge. The weak-field linearized theory of general relativity unveils a mathematical structure comparable to the Maxwell equations and it splits gravitation into components similar to the electric and magnetic field. Moving charge creates a magnetic field according to Ampere's law. Analogously, moving mass creates a mass current which generates a gravitomagnetic field. For the light ray grazing the limb of the Sun, the gravitomagnetic time-delay is smaller than the effect produced by the mass of the Sun by terms of order $\omega L/c$, where $\omega$ is the Sun's angular velocity of rotation and $L$ is its radius. The gravitomagentic effect (Lense-Thirring effect) due to the solar rotation in the time delay is about 20 ps; this is comparable to that due to solar quadrupole moment. Lense-Thirring time-delay of light can be separated from the time-delay caused by the mass and quadrupole moment due to their different signitures. With 3 ps accuracy in a normal point in range, this effect can be measured to 10 %. Since the Lense-Thirring effect due to Earth is already verified to 10 % [22], this gives a solar angular momentum measurement to 10 % which is useful as a consistency check with solar modeling value based on solar seismology

Possible variation of fundamental constants is a focus point in cosmology. Among fundamental constants, the Newtonian gravitational constant G is most directly related to cosmology. ASTROD I will be able to determine its variation with an order of magnitude improvement from previous efforts.

Concerning solar system tests of modified gravity theories, in relation to dark matter and dark energy, we mention that the MOND theory [23, 24] and anomalous Pioneer acceleration [15] could be measured and tested to very high accuracy. During the Venus flyby, the 'flyby anomaly' [25] could also be measured and tested to very high accuracy.

A summary of these goals is compiled in Table 2.

**Table 2** Summary of the scientific objectives of the ASTROD I mission

| Effect/Quantity | Present accuracy [16] | Projected accuracy |
|---|---|---|
| PPN parameter $\beta$ | $2 \times 10^{-4}$ | $3 \times 10^{-8}$ |
| PPN parameter $\gamma$ (Eddington parameter) | $4.4 \times 10^{-5}$ | $3 \times 10^{-8}$ |
| (dG/dt)/G | $10^{-12}$ yr$^{-1}$ | $3 \times 10^{-14}$ yr$^{-1}$ |
| Anomalous Pioneer acceleration $A_a$ | $(8.74\pm1.33)\times10^{-10}$ m/s$^2$ | $0.7 \times 10^{-16}$ m/s$^2$ |
| Determination of solar quadrupole moment parameter $J_2$ | $1 - 3 \times 10^{-7}$ | $1 \times 10^{-9}$ |
| Measuring solar angular momentum via solar Lense-Thirring Effect | 0.1 | 0.1 |
| Determination of planetary masses and orbit parameters | (depends on object) | 1 - 3 orders better |
| Determination of asteroid masses and density | (depends on object) | 2 - 3 orders better |

These goals are based on the orbit simulation for an orbit suitable for launch in 2012 with timing uncertainty 10 ps, and drag-free uncertainty is $10^{-13}$ ms$^{-2}$Hz$^{-1/2}$ at 0.1 mHz [26]. The simulation is from 350 days after launch for 400 days with 5 observations per day. In the numerical integration of for this orbit simulation, we have taken the following effects into account:

(i) Newtonian and post-Newtonian point mass gravitational interaction between every two bodies including the Sun, nine planets, Moon, three big asteroids (Ceres, Pallas and Vesta)
(ii) Newtonian attraction between a body with gravitational multipoles including Sun (J2), Earth (J2, J3, J4) and Venus (J2), and others as point masses
(iii) the spacecraft is treated as a test body in the numerical integration.

Other similar choices of orbit have similar results.

In the present proposal, our timing uncertainty is 3 ps and drag-free uncertainty is $3 \times 10^{-14}$ m s$^{-2}$ Hz$^{-1/2}$ at 0.1 mHz. These uncertainties are 3.3 times lower than the uncertainty figures used in the orbit simulation of Ref. [26]. Since for precision orbit, the simulation is highly linear, we can safely divide the simulated uncertainties of parameters by 3.3 to obtain the uncertainties for the present proposal. The results are shown in Table 2 with the exception of the solar Lense-Thirring effect which was estimated and explained in this section. When the observation times in the orbit simulation is decreased to one time per day instead of five times per day, the uncertainty increased by about 10 per cent. A more thorough orbit simulation including two Venus flybys (and the period after injection orbit correction is underway.

*ASTROD I would be a start of a new era of probing the fundamental laws of spacetime and mapping solar-system of gravity together with fixing the dynamical reference frame and its links. In the second phase, 3-spacecraft ASTROD (ASTROD II) would be envisaged to test general relativity to 1 part per billion, to detect solar g-modes, to measure the solar Lense-Thirring effect to 10 ppm, and to probe gravitational-waves at frequencies below the LISA bandwidth* [1, 2, 4]. *In the third phase (Super-ASTROD [ASTROD III]), larger orbits can be implemented to map the outer solar system, and to probe primordial gravitational-waves with even lower frequencies* [27].

The objectives and method of measuring solar Lense-Thirring effect for ASTROD have been stated and analyzed in [1, 2, 4]. The objectives of gravitational-wave detection have been discussed in [1, 2, 4] also. Since the solar g-mode detection is of wide interest in solar-physics community and is one of the main motivations of next stage mission after ASTROD I, we discuss this objective in following subsection.

## 2.1 The internal structure and dynamics of the Sun [28]

For the past 20 years, traditional helioseismology techniques have provided a wealth of information regarding the internal structure and dynamics of the Sun in both the convection and radiation zones [29]. Probing the Sun's interior, from the solar surface to depths of approximately 0.2 solar radii, has been achieved by direct analysis of the pressure modes detected in Doppler observations of surface radial velocities.

However, the gravity modes (g-modes), that probe the very core of the Sun where thermonuclear reactions take place (approximately less than 0.2 solar radii), remain elusive. These modes have very small amplitudes and generate very small radial velocities, thereby making them very difficult to detect using conventional techniques. There have been several unsubstantiated attempts at detecting g modes using ground- and space-based instrumentation ([30] and references therein). Furthermore, Garcia et al. [31] recently claimed to have detected a collective signature of the dipole g modes, but this is also yet to be confirmed.

In order to make significant progress towards a *total* understanding of the internal structure and dynamics of the Sun, from the solar surface down through the core to the very center, we require an unambiguous detection of solar g-modes. ASTROD (ASTROD II) [2, 32, 33] will deliver a new unconventional method [2, 33-36 and references therein] that is capable of achieving this objective.

Fig. 3 displays various, theoretical and observed, solar velocity amplitudes of the g-modes, as well as the anticipated sensitivities for the space-instruments, ASTROD and LISA [28]. The two theoretical models comprise the most pessimistic estimate from Kumar et al. [37] and the most optimistic estimate from Gough [38] (also see [36] for a good analysis) and the best up-to-date observational limits come from [31]. It is immediately clear that the gravitational wave detector, LISA, is very unlikely to detect g-modes. However, by assuming the best possible gravity-strain sensitivity for ASTROD ($10^{-23}$ at 100 μHz) and even the most pessimistic amplitudes of Kumar et al., *ASTROD will deliver unambiguous detection for modes with frequencies between 100 μHz and 300 μHz. Moreover, the detection of such high frequency g-modes will provide a better diagnostic of the inside of the Sun than any lower frequency g-mode (below 100 μHz)*. We refer the readers to ref.'s [28, 33, 36] for details.

## 3. Mission Profile

### 3.1 Orbit requirements

The spacecraft needs to reach the far side of the Sun to measure the Shapiro effect (relativistic light retardation) to high precision. A trade-off is needed between minimising the time to reach the far side of the Sun because of the risk to the mission from aging of equipment, and reducing the amount of fuel needed for the mission. If the spacecraft is launched into solar orbit with a period about 300 days at a suitable epoch, the perihelion can reach 0.72 AU to encounter Venus. If we tune the Venus encounter such that the spacecraft changed into a 225-day orbit, the spacecraft has a chance to have a second encounter with the Venus one-period or half a period later. This time we design the orbit to have a good swing toward the Sun. This way the spacecraft will reach the other side of the Sun in about 370 days. Our design concept saves fuel and the spacecraft still reaches the other side of the Sun quickly. The science observation starts within a couple of months from launch after the orbit injection is done. Drag-free operation will be operative through the whole science observation.

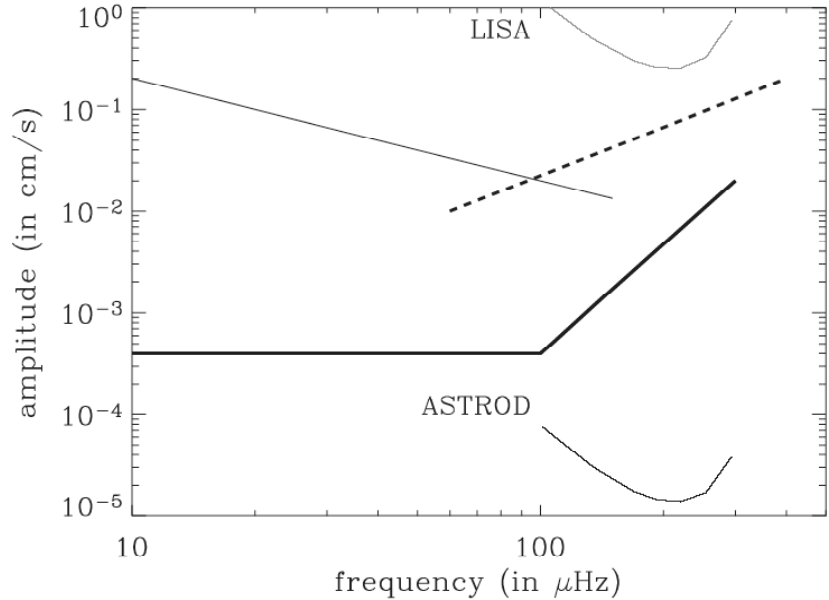

**Fig. 3** Comparison of surface radial velocity amplitudes for l=2 g-modes (quadrupole modes) (explanation in the text below). Theoretical estimates from [37] (thick solid line) and [38] (thick dashed line); present observation one sigma limit corresponding to an average of 50 modes observed by the GOLF instrument with 10 years of data, derived from [31] (thin straight line); LISA one sigma limit (grey solid line) assuming a one-year integration time and a strain sensitivity of $10^{-23}$ at 3000 μHz and a $f^{1.75}$ dependence [10]; ASTROD one sigma limit (thin solid line) assuming a one year integration time and a strain sensitivity of $10^{-23}$ at 100 μHz and a $f^2$ dependence [4, 39], with a inner spacecraft perihelion at 0.4 AU and an outer spacecraft aphelion at 1.32 AU. The surface velocity amplitudes for ASTROD were derived using the most recent gravity strain sensitivities in [4, 39] and the equations in [35]. The gravity strain falls off as $1/R^4$, $R$ distance to the Sun. The significant improvement provided by ASTROD with respect to LISA is mainly due to orbit configuration to the Sun and better strain sensitivity.

Earth's orbit (1 AU) around the Sun has a period of 365 days. Venus' orbit (semimajor axis 0.723 AU) around the Sun has a period of 225 days. The Venus' synodic period is 584 days. For a launch of a spacecraft into a solar orbit with aphelion around 1 AU (Earth) and perihelion around 0.723 AU (Venus), the semi-major axis is 0.808 AU. By Kepler's law, the period of this orbit is 294 days and it takes 147 days to reach the Venus. At the launch position, Earth should be ahead of Venus by 55° so that Venus could catch the spacecraft just in time. Around January, 2007, Earth and Venus are in such relative positions and launches are possible. With these simple calculations, first Venus encounter would be around 150 days after launch; second Venus encounter would be around either 260 days after launch or 370 days after launch. Because of the eccentricity and inclination of Earth and Venus orbits, and the small but nonzero angles of encounter, actual numbers have a range. However, these numbers serve as starting points for orbit design.

### 3.2 Launcher requirements

The ASTROD I spacecraft will first be launched to a Transfer Orbit of large eccentricity with a perigee of between 200 km and 300 km and an apogee of about 500,000 km. The propulsion to solar orbit with an orbital period of 300 days will be achieved using a 100 kg liquid propulsion unit, which will be separated after its usage. The injection correction will be performed using a medium-sized ion thruster on board ASTROD I.

There are two Chinese launchers CZ-4B and CZ-2C, which can send the spacecraft into a large eccentricity orbit with 200 × 500,000 km. These 2 launchers are good candidates for ASTROD I with a total mass of about 490 kg.

### 3.3 Ground segment requirements

The ground segment is composed of a laser ranging station with a special instrumentation to measure accurately both the start and return time of laser pulses. There are roughly 40 laser stations in the world that range satellites and moon regularly. A typical laser station includes a pulsed Nd:YAG Laser (20 to 200 ps) at 10 Hz, a telescope (~ 1 m diameter), 2 event timers or an intervalometer. For the ASTROD I mission, event timers are required to proceed absolute measurements into the TAI time scale. For ASTROD I, the requirements on pointing and tracking capability are both better than 1 arcsec. Adaptive optics is assumed. These can be reached with the 1.5 m telescope CERGA at the Côte d'Azur station and with the 1.2 m telescope at Yunnan station.

Furthermore there have to be deep space stations available for RF communication and tracking. The stations must possess X-Band capability.

### 3.4 Drag-free system requirement

To achieve its scientific objectives, ASTROD I must implement drag-free control. The acceleration noise spectral density that ASTROD I requires for 2 orbit plane axes is

$$S_{\Delta a}^{1/2}(f) = 10^{-14} [(0.3 \text{ mHz}/f + 30 \times (f/3 \text{ mHz})^2] \text{ m s}^{-2} \text{ Hz}^{-1/2}, \qquad (2)$$

for 0.1 mHz ≤ $f$ ≤ 0.1 Hz where the frequency, $f$, is given in units of mHz. For perpendicular axis, this condition can be relaxed by one order of magnitude.

In equation (2) it has been assumed a factor 3.3 improvement with respect to the ASTROD I requirement considered in [41]. As explained in section 1, the tightening of this requirement is to support the tightening by a factor of 3.3 of the timing accuracy and is deemed feasible in the time frame of Cosmic Vision period from the experience of GOCE sensor heads. At 0.1 mHz an acceleration noise spectral density of 3 × $10^{-14}$ m s$^{-2}$ Hz$^{-1/2}$ is then required. The most stringent acceleration noise required is about 1.3 × $10^{-14}$ m s$^{-2}$ Hz$^{-1/2}$ at approximately 0.36 mHz. Figure 3 shows ASTROD I acceleration spectral noise density target given by equation (2) compared to LISA [10] and LISA Pathfinder LTP (LISA Technology Package) [11].

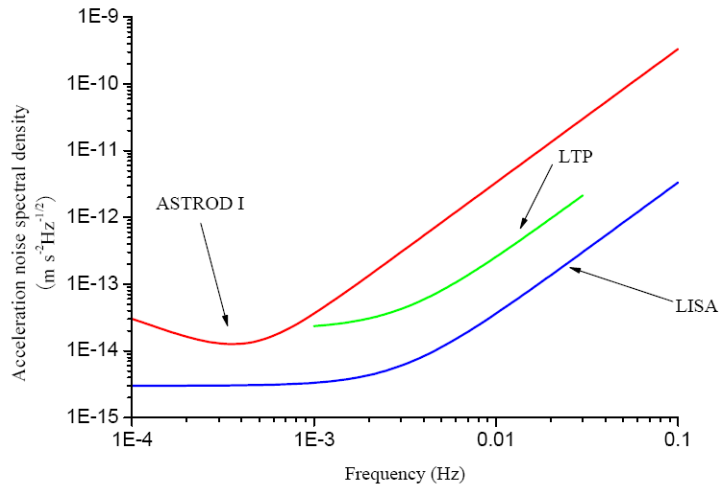

**Fig. 4** A comparison of the target acceleration noise curves of ASTROD I, the LTP and LISA

The total acceleration noise disturbance $f_p$ acting on the proof mass (PM) in the control loop can be written as [40, 41]

$$f_p \approx X_{nr}(-K) + f_{np} + (f_{ns} + T N_t) K \omega^{-2} u^{-1}, \qquad (3)$$

where $X_{nr}$ is the sensor readout disturbance, $K$ proof mass-spacecraft coupling stiffness (spring constant), $f_{np}$ environmental disturbance on the proof mass, $f_{ns}$ external environmental disturbances on the spacecraft, $T$ is a thruster factor which is nominally 1 whose effects can be absorbed into thruster noise disturbance $N_t$, $\omega$ the circular frequency $2\pi f$, and $u$ the control loop gain.

The first term in equation (3) represents the contribution of the sensor readout disturbance, $X_{nr}$. The second term, $f_{np}$, is the contribution due to total direct acceleration disturbances acting on the proof mass. Direct acceleration disturbances are themselves divided into environmental and sensor back action disturbances. Finally, the last term represents the contribution of the external disturbances acting on the spacecraft, i.e. solar radiation pressure, solar wind, thruster noise, etc, that couples to the proof mass motion because of residual proof mass-spacecraft stiffness, $K$. This term will be suppressed via a high control loop gain, $u$. These disturbances are discussed in detail in [41] for ASTROD I.

In the drag-free system requirement, we consider a root square sum (rss) of those three terms to obtain the total acceleration noise contribution. We also attribute to each of them an equal portion of the noise budget. The acceleration noise spectral density, at 0.1 mHz, should then be less than $1.7 \times 10^{-14}$ m s$^{-2}$ Hz$^{-1/2}$ for each term on the right of equation (3).

The parameter values used for the disturbance estimations are given in Table 1 of reference [41]; all the values are the same except patch-field needs to be compensated to 0.01 V. Estimates of acceleration disturbances directly acting on the proof mass (magnetic, thermal, impacts, etc) and sensor back-action disturbances on the proof mass are given in Table 3 of reference [41]; all the values are the same except that the patch-field PM sensor back-action acceleration disturbances are decreased by tenfold to $2.9 \times 10^{-15}$ m s$^{-2}$ Hz$^{-1/2}$ and patch field needs to be compensated to 0.01 V; the total rss value of acceleration disturbances is $8.6 \times 10^{-15}$ m s$^{-2}$ Hz$^{-1/2}$, which is below the requirement specified above. *The important change from earlier version of ASTROD I is that the patch field needs to be measured and compensated to 0.01 V.* As a general note, the LISA study team is considering the possibility of measuring and compensating voltage difference across capacitors to considerably better than 0.01 V [42, 43].

The acceleration noise target for the other two terms forces us to require certain values for the sensor readout noise, $X_{nr}$, and drag-free control loop gain, $u$. At 0.1 mHz, and assuming a stiffness of $3.1 \times 10^{-7}$ s$^{-2}$ [41], we will require $5.6 \times 10^{-8}$ m Hz$^{-1/2}$ as the sensor readout noise. In the case of disturbances caused by external forces acting on the spacecraft the major contribution will be of thruster noise. We have considered a thruster noise level of 10 μN Hz$^{-1/2}$. If that is the case, at 0.1 mHz, we need to achieve control loop gains larger than $1.3 \times 10^6$.

The most stringent requirement on acceleration noise spectral density is at a frequency of approximately 0.36 mHz. At that frequency and under the same assumptions discussed above, the different terms of equation (3) should contribute less than $7.3 \times 10^{-15}$ m s$^{-2}$ Hz$^{-1/2}$. Taking into account the frequency dependence of the different direct acceleration disturbances (see [39]) and their values at 0.1 mHz, it can be shown that they accomplish this requirement. For the other two terms a sensor readout noise of approximately $2.3 \times 10^{-8}$ m Hz$^{-1/2}$, and a control loop gain larger than $2.4 \times 10^5$, which is less stringent than the one quote at 0.1 mHz, are required.

The acceleration disturbances and requirements for ASTROD I at 0.1 mHz are finally summarized in Table 3.

### 3.5 Thermal requirement

One of the concerns for ASTROD I mission is the thermal fluctuating environment associated to the changing spacecraft-Sun relative distance along its orbit. Spacecraft-Sun relative distance will vary from approximately 0.5 AU to 1.04 AU. In that case, the total solar irradiance flux will vary by a factor of 4 from the closest to the far off spacecraft distance to the Sun.
ASTROD I mission baseline configuration is a two one-way laser pulse ranging. Since the precision in ranging is 0.9 mm. It is much less stringent than that of LISA or ASTROD. The drag-free requirement is also less. In this case, the absolute thermal requirement for the inertial sensor is less stringent compared to, for example, what LISA or ASTROD requires. The telescope also requires a less stringent level of thermal stability for proper functioning. The spacecraft absolute temperature needs to be controlled to $\Delta T = \pm 1$ K.

The changes in the distance between the ASTROD I spacecraft and the Sun might force us to use not only passive thermal isolation techniques but also active thermal control. Ideas like changing area of the radiative panel, either by moving shutters and/or by changing the orientation of certain components, and phase changing materials for active thermal control and heat rejection and storage, need to be considered. Active thermal control by using moving components might not be compatible with the ASTROD I drag free stringent requirement. Nevertheless all these aspects need to be carefully addressed in the assessment study.

**Table 3** Acceleration disturbances (rss amplitudes) and requirements for ASTROD I.

| | ASTROD I |
|---|---|
| (a) Estimated total acceleration disturbance (at 0.1 mHz): | |
| $f_p$ [m s$^{-2}$ Hz$^{-1/2}$] | $2.6 \times 10^{-14}$ |
| (b) Estimated contributions to $f_p$ (at 0.1 mHz): | |
| $f_p \approx X_{nr}(-K) + f_{np} + (f_{ns} + TN_t)K\omega^{-2}u^{-1}$ | |
| Direct PM acceleration disturbances: | |
| $f_{np}$ [m s$^{-2}$ Hz$^{-1/2}$] | $8.6 \times 10^{-15}$ |
| Direct SC acceleration disturbances: | |
| $f_{ns} + TN_t$ [m s$^{-2}$ Hz$^{-1/2}$] | $2.8 \times 10^{-8}$ |
| PM-SC spring constant: | |
| $K$ [s$^{-2}$] | $3.1 \times 10^{-7}$ |
| (c) Inferred requirements | |
| Position readout noise: $X_{nr}$ [m Hz$^{-1/2}$] | $2.3 \times 10^{-8}$ |
| Control-loop gain: $u$ | $1.3 \times 10^6$ |

Passive thermal isolation can be implemented by choosing materials with a low coefficient of thermal expansion to minimize gravitational instabilities in the proof mass region, and low emissivity to minimize heat radiation coupling. In the internal layers, temperature gradients have to be minimized to suppress radiative coupling between the structure and science payload. A thermal shield for the optical bench, similar to LISA, with gold coated surfaces can minimize radiative coupling [10]. Temperature instabilities due to power driven payload components (electronic boxes, etc) can also couple to the thermal environment of the proof mass. Thermal fluctuations generate unwanted proof mass acceleration disturbances that need to be suppressed to a certain required level (see Table 4 of [41]). Laser light will not impinge directly onto the proof mass but in a fiducial point of the spacecraft. Relative position of the proof mass and the spacecraft fiducial point can change because of thermal expansion and needs to be monitored; this will not be serious since our accuracy is 3 ps (0.9 mm). All these effects need to be carefully accounted for.

A preliminary requirement on the proof mass housing thermal stability has been established to be below $10^{-5}$ K Hz$^{-1/2}$ in the measurement bandwidth $0.1$ mHz $\leq f \leq 0.1$ Hz [41].

Thermoelastic fluctuations due to the different thermal environmental stages along the orbit induce changes in the self-gravitational configuration of the spacecraft structure and payload. These effects can cause a drift in the self-gravitational force and unwanted gravitational fluctuating forces acting on the proof mass. Spacecraft and payload materials, thermal properties (thermal conductivity, coefficient of thermal expansion, emissivity, etc.), and assembly (geometry and distance to the proof mass) have to be carefully considered to quantify these effects.

### 3.6 Charge Management requirement

One of the main sources of acceleration disturbances is due to charge accrued on the proof mass. As ASTROD I cruises through space, its test mass will accrue charge due to galactic cosmic-rays (GCRs) and solar energetic particles (SEPs) incident on the spacecraft. The resulting total charge accrued and charge fluctuations can cause acceleration disturbances because of Lorentz forces, due to interaction with the interplanetary magnetic field, and Coulomb forces, and due to interaction with the conducting surfaces of the capacitive sensor and housing. *For ASTROD I disturbance estimation due to charge accrued and charge fluctuations we have required a maximum total proof mass charge accrued of $q = 10^{-12}$ C.* A fluctuating charge of $\delta q = 6.1 \times 10^{-15}$ C Hz$^{-1/2}$ at 0.1 mHz has been considered and used in the estimates.

Spurious coherent signals will also arise because of the accumulation of test-mass charge over time. In a preliminary work using the GEANT4 toolkit [44, 45], we simulated the charging process due to GCRs and SEPs [46, 47]. We predicted a total net charging rate about 33.1 +e/s from cosmic-rays between 0.1 and 1000 GeV at solar maximum, and 73.1 +e/s at solar minimum with uncertainty of ±30% [46]. We also obtained charging rates due to SEPs ranging from 2840 to 64300 +e/s at peak intensity, for the four largest SEP events that occur in September and October 1989 [47]. The implementation of a charging management system is discussed in section 4 on the payload instruments.

### 3.7 Sunlight shielding requirement

The science observation of ASTROD I starts a couple of months after launch until the end of mission life time (3-8 years). During day time and during the Shapiro time-delay measurement (when the spacecraft is on the other side of the Sun), the sky light and sunlight will shine directly into the optical components for laser signal detection if there is no shielding system. When the spacecraft is nearly opposite to the Sun, the background light shining will then be of the order of 100-400 W (in the infrared and visible electromagnetic spectrum region). The receiving laser light is of the order of 100-400 fW. In order to distinguish laser light information (science signal) from sunlight irradiation, a rejection factor of $10^{-18}$ of sunlight power reaching the photo detector should be required. This makes received sunlight power at about 0.1 % compared to received laser light at the detector. This requirement is met by three filterings:
(i) Spatial filtering: $10^{-9}$,
(ii) Spectral filtering: $10^{-3}$;
(iii) Temporal filtering: $10^{-6}$.

The ASTROD I *sunlight shield system will essentially consist of a geometrical shutter (pinhole), dielectric optical filters and a time gate to meet the requirement.* Design detailsn will be discussed in section 4.

### 3.8 Other requirements and issues

For drag-free system, one requirement is on the reliability and lifetime of μ-Newton thrusters for drag-free optimum performance. The contribution to the total proof mass acceleration spectral density of thruster noise, $TN_t$, for a 1.75 kg proof mass is, in units m s$^{-2}$ Hz$^{-1/2}$, approximately

$$f_p^{th} \approx 1.7 \times 10^{-14} \left( \frac{TN_t}{10\ \mu\mathrm{N\ Hz}^{-1/2}} \right) \left( \frac{1.75\ \mathrm{kg}}{m_p} \right) \left( \frac{K}{3.1 \times 10^{-7}\ \mathrm{s}^{-2}} \right) \left( \frac{1.3 \times 10^6}{u} \right) \left( \frac{10^{-4}\ \mathrm{Hz}}{f} \right)^2 \quad (4)$$

where $K$ denotes the proof mass-spacecraft coupling, $f$ is the frequency, $m_p$ is the mass of the proof mass, and $u$ is the control loop gain. If thrusters of force noise of 10 μN Hz$^{-1/2}$ with enough reliability were not available we could consider the case of thrusters with noise of the orders of 0.1-1 mN Hz$^{-1/2}$. The acceleration noise spectral density due to thruster noise would then be approximately $1.7 \times 10^{-13}$ ($10^{-12}$) m s$^{-2}$ Hz$^{-1/2}$ at 0.1 mHz for a given 0.1 (1) mN Hz$^{-1/2}$ thruster noise. To accomplish the ASTROD I drag-free requirements the control loop gain should then be improved by one (two) orders of magnitude.

Of special concern are issues like the longevity and reliability of technologies, like inertial sensor, μ-Newton thrusters, lasers (4 on board of ASTROD I spacecraft) and diverse optical components in the space environment. A preliminary environmental study for ASTROD I has been performed by the DESSTD (Deep Space Exploration & Space Science Technology Division) set up by CAST (Chinese Academy of Space Technology). The ASTROD I spacecraft will vary its relative distance to the Sun (from 0.5 AU to 1.04 AU) and will be out of the Earth magnetosphere. Galactic cosmic rays (GCR), solar energetic particles (SEP), solar wind and solar electromagnetic radiation are some of the environmental hazards ASTROD I will encounter. Galactic cosmic rays are very energetic particles, with energies of the order of GeV, isotropically distributed and found everywhere in the interplanetary space. GCR are mostly ionized having cyclical variations in flux levels. Galactic cosmic ray flux increases with distance to the sun. The magnitude of galactic cosmic ray gradients varies from 10 % per AU at 1 AU to 4 % per AU at 5 AU. Being conservative we could consider evaluating hazard with GCR data at 1 AU.

Solar energetic particle events dramatically increase proton and heavy ions fluxes. Proton energies will be of the order of hundreds of MeV while heavier ions will carry energies of the order of hundreds of GeV. Flux abundance depends on the distance to the Sun, d. Peak intensity of solar energetic particles scales with Sun distance as $1/d^3$, being 8 times more intense at 0.5 AU than at 1 AU while the total fluence scales with $1/d^2$. A solar wind spectrum is not well known. How does the solar wind affect the spacecraft? It produces an acceleration and acceleration noise directly acting on the spacecraft comparable to solar electromagnetic radiation. A particle monitor can be considered as a payload to monitor environment.

In the case of solar electromagnetic radiation solar irradiance is approximately 1380 W m$^{-2}$ at 1 AU. The effects of solar electromagnetic radiation on the spacecraft are numerous. Risks due to total radiation dose effects, displacement damage, single event effects, optical darkening and thermal property changes due to solar radiation exposure, and optical contamination have to be considered. Single event effects would be a major concern for ASTROD I during intense solar energetic particle events. Also a detailed study to evaluate the effects of spacecraft material exposed to radiation, like changes in thermal properties of spacecraft materials should be performed.

**4. Proposed Payload Instruments**

In this section, we discuss payload instruments. A schematic diagram is shown in Fig. 1. A Cassegrain telescope transmits and receives laser pulses. The received laser pulses go through pinhole and dielectric filter to reach the optical bench and are separated into two paths according to their wavelengths to be detected and timed by photo detectors and an event timer. The emitting light is also timed by the event timer using part of the beam. The emitting light pulses go through the optical bench and the telescope to be transmitted to the Earth laser ranging stations and the arrival times recorded. Transmission times of laser pulses from the Earth laser ranging stations are also recorded. The Cesium clock provides the time and frequency standard. The synchronization procedure with the ground clock is the same as that of T2L2 experiment [8]. The transmission and receiving of laser pulses are asynchronous. The ranges are deduced from the receiving and transmitting times of this two one-way two-wavelength ranging. The drag-free system (including inertial sensor, thermal control and diagnostics, and μN thrusters) is for minimizing non-gravitational disturbances to a level supporting the same range accuracy in a sufficient long period of time (i.e., a year or so). A passive payload shield is foreseen to shield the sensitive components from the space environment. Its mass is 15 kg.

In the following subsections, we discuss various payload instruments. Table 4 compiles the mass, power, and size budgets of the payload instruments.

**Table 4 Payload budgets**

| Component | Mass [kg] | Power [W] | Size [mm] |
|---|---|---|---|
| **Cassegrain telescope** | 25 | - | ⌀ 400×600 |
| **Optical bench** | 5 | 2 | tbd |
| **Nd:YAG pulsed-laser (4)** | 32 | - | Each (4x) 100×100×100 |
| **Laser power supply** | 2 | 45 | 100×100×100 |
| **Cesium clock** | 15 | 33 | 160×200×400 |
| **Laser pulse timing system (2)** | 16 | 80 | 280×230×180 |
| **Inertial sensor** | 6 | 10 | 300×300×300 |
| **Thermal control system** | ≈ 10 | tbd | tbd |
| **μN-Thruster assemblies** | 22 | 100 | tbd |
| **Charge management system** | ≈ 4 | 2 | 50×50×100 |
| **Payload shield** | 15 | - | tbd |
| **Total** | 160 | 280 | |

## 4.1 Cassegrain telescope

An optical telescope (Fig. 5) with a diameter of 300 mm is needed for both transmitting and receiving the laser signals from and to the ground station. In order to ensure a small size (length) of the telescope a f/1-system (f-ratio of the primary mirror) is foreseen. The secondary mirror then creates a convergent light beam. The focal ratio for the whole Cassegrain system is f/10. The equivalent focal length (EFL) for a f/10-sytem is 3000 mm.

Behind the telescope a shielding system takes place that reduces the amount of sunlight passing into the optical bench that follows the shielding system. Especially when the line of sight (from the spacecraft to the earth) is near the solar limb, an effective shielding system is important. A pinhole plate is installed in the focal plane of the telescope for this purpose. The adjustable pinhole plate can be actuated electromechanically for adjustment. Only a fraction less than $10^{-9}$ of the 100-400 W sunlight power, i.e., 100-400 nW, enters the part after the pinhole because in this configuration the Sun and most part of the corona are out of sight. At sunspot maximum the total coronal light beyond 1.03 solar radius is $1.5 \times 10^{-6}$ solar flux; at sunspot minimum $0.6 \times 10^{-6}$ solar flux. With 1 mm pinhole, we only see less than $10^{-3}$ of solar corona. Hence, pinhole attenuation factor is $10^{-9}$. The pointing accuracy has to be better than 0.575 arcmin in order to keep the object (the earth) in the field of view. This should be easily implemented.

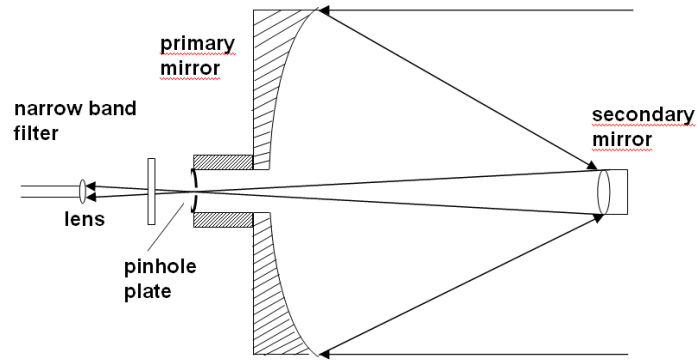

**Fig. 5** Scheme of the Telescope

The pinhole plate is coated with a reflecting coating, so that *all the sunlight that does not pass through the pinhole is reflected back to space*.

A 1.0 mm pinhole would not create significant diffraction. Behind the focus point a narrow band filter (0.5 nm bandwidth) reflects all light except light with the wavelengths of the laser light (532 nm and 1064 nm). As the narrow band filter is placed behind the pinhole the filter does not affect the back reflection of the residual coronal sunlight. The light power behind the pinhole would be less than $10^{-9} \times 400$ W = 400 nW, which is no problem for the filter

The image scale at the focal plane is s = f/206.265 [mm/arcsec], where f is in meters. Assuming an equivalent focal length of 3 m (f/10-system), the angle of view of a 1.0 mm pinhole therefore is 68.755 arcsec which is 1.15 arcmin. For an angular distance of several arc minutes between the earth ground station and the solar limb viewed from the spacecraft, this concept will prevent direct sunlight to come into the optical bench. The front side of the pinhole plate is coated with a reflecting coating so that the wasted light is reflected back to space through the telescope. Behind the pinhole a lens transforms the divergent light beam into a parallel one.

The mass of a weight-optimized system is 25 kg including the secondary mirror, the shielding system and the mounting system. The point-ahead angle for the telescope is about 20 arcsec (varying) and is different for incoming and outgoing light. Point-ahead mechanism needs to be implemented. However, since direct photo detection is used and the pinhole has a 68 arcsec field of view, This can be implemented on changing the angle of outgoing beam.

## 4.2 Pulsed laser systems and the measurement concept

Pulsed laser beams are being sent in both directions between ground station (say, a 1.2 m telescope) and the ASTROD I spacecraft. This two one-way laser ranging concept leads to a much better performance compared to single one-way ranging concepts. To measure and subtract out the effect of the air column of the atmosphere, the ranging is done with two sufficiently different wavelengths $\lambda_1$ and $\lambda_2$. Recently, a more accurate atmospheric model [49, 50] has been proposed and used by the satellite laser ranging and lunar laser ranging community. The accuracy of this model is confirmed to a couple of mm's. Since the optical pathlength is wavelength dependent, with two wavelengths and atmospheric model, the actual atmospheric delay can be measured using their timing difference to sub-millimetre accuracy. Interplanetary refraction is small. The choice of these wavelengths will be determined by the availability of lasers with sufficiently high pulse powers, where for the lasers onboard the spacecraft mean powers of the order of 1-2 W are the baseline (two plus two spare pulsed Nd:YAG lasers). Possible candidate wavelengths are $\lambda_1$ = 1064 nm and $\lambda_2$ = 532 nm. The pulses will come at a repetition rate of about 100 Hz, a pulse width of 50 ps and 10 mJ pulse energy, but this may be subject to laser and other considerations. To reach 3 ps accuracy, we need 280 photons of combined observations. A bunch of consecutive observations can be grouped together into a normal point using a simple orbit model. The ground laser could be similar or with slightly larger power. Recently, station positions have reached a couple of millimetre accuracy. We do expect, in the time-frame of ASTROD I mission, these station positions will reach millimetre accuracy.

### 4.3 Optical bench

The function of the optical bench (Fig. 6) is to process the light collected by the telescope and to generate information to be sent to the ground station. These tasks must be fulfilled under the boundary condition of keeping the influx of sunlight to a negligible amount. Figure 6 shows a schematic diagram.

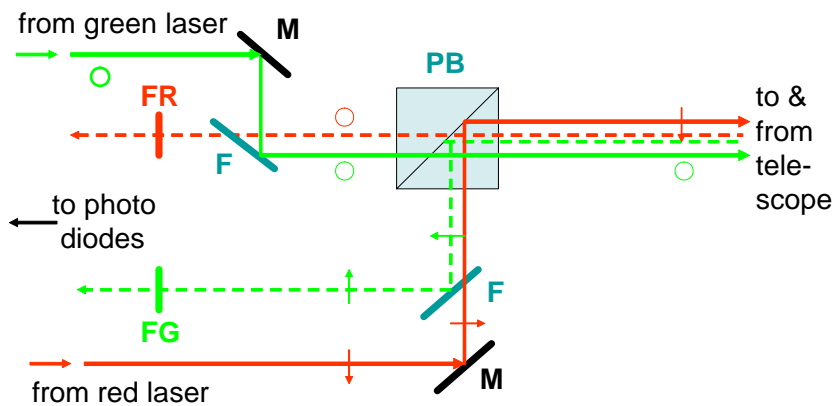

**Fig. 6** A schematic diagram of the optical bench (PB: **polarizing beamsplitter,** M: **mirror,** F: **dielectric filter,** FR: **filter red,** FG: **filter green**)

#### 4.3.1 Light paths of the two incoming laser beams

The common use of the telescope for the four laser beams is sketched in Fig. 6. The (low power) beams from the ground station are shown as dotted lines, and in the arbitrary colours of red and green. They come in with crossed polarizations: red with polarization out of the drawing plane (symbolized by the circles), and green with polarization inside the plane (arrows).

A polarizing beam splitter (PB) separates these beams, letting the red beam pass, and reflecting the green beam (downward). The two identical multilayer dielectric filters (F) are designed to transmit the red light and reflect the green. Finally the red and green beams pass through appropriately designed dielectric filters (FR and FG, respectively) for further reduction of the sunlight, by factors given by the bandwidth of these filters, typically of 0.5 nm or less, giving further sunlight suppression by a factor 1000. If needed, a high-finesse Fabry-Perot in each of the paths could improve the filtering. Polarization alignment and interplanetary depolarization due to dust scattering to 1 % will not create a problem for this optics scheme and detection accuracy.

The two incoming beams, still with their original polarizations (normal for red, in-plane for green) are now sent to the avalanche photodiode (APD) arrays (4×4 or 5×5) to monitor the pulse arrival times. These photodiode arrays have the additional task of aligning the spacecraft to the incoming beams.

### 4.3.2 Light paths of the two outgoing laser beams

The same telescope is used for the pulsed laser beams emitted by the two lasers onboard the spacecraft. Their wavelengths are essential identical to those of the two incoming beams. The outgoing beams have polarizations opposite to the incoming ones of the same colour. Thus the (heavy, solid) green beam (circle) is transmitted by the polarizing beam splitter PB, whereas the (heavy, solid) red beam (arrow) is reflected, and thus both exit via the telescope.

### 4.3.3 Received laser light

For the measurement of the post-Newtonian parameter $\gamma$ (Eddington light-deflection parameter) near the opposition, i.e., 1.7 AU = 255 million km away from Earth, the spacecraft telescope will then pick up only about $10^{-13}$ of the power emitted. For the wavelength of 1064 nm that would mean a pulse of $10^{-15}$ J or 5,000 photons. Since we are using APD array for single photon detection, we may needs to attenuate the beam before the it enters the detector.

At the same time, the sunlight will, at opposition, shine with 400 W into the spacecraft telescope (see Section 3.7) and extreme care must be taken to reduce that sunlight to a level that the laser signal is dominant. With the pinhole shutter, narrow bandwidth dielectric filters and time-filtering (10 ns window) of the laser pulse, the remaining solar photons will be sufficiently less than the laser photons at the photo detector and laser pulse signals can be detected readily.

### 4.3.4 Advantages of the setup

The set-up has several very distinct advantages:
- it is made up of standard, off-the-shelf components
- the beams are, by their polarization, well distinguished from one another
- the incoming beams can be separately filtered with standard dielectric filters
- the incoming and outgoing beams of identical color have opposite polarization, and thus low crosstalk
- any crosstalk of ingoing and outgoing beams can additionally be reduced by temporal filtering

## 4.4 Cesium clock

There are two methods for laser ranging between spacecraft and ground laser stations. One method is interferometric ranging similar to LISA, the other method is pulse ranging similar to Lunar Laser Ranging (LLR). For ASTROD I the pulsed laser ranging will be utilised. The emitting times and receiving times will be recorded by a cesium atomic clock. For a ranging uncertainty of 0.9 mm in a distance of $2.55 \times 10^{11}$ m (1.7 AU), the laser/clock frequency needs to be known to one part in $10^{14}$ by comparison with ground clocks over a period of time. Stability to $6 \times 10^{-14}$ in 1000 s (round-trip time) is required.

The company Symmetricom offers a space qualified Cs-clock (4415 – Digital Cesium Frequency Standard) with this required stability. Symmetricom's 4415™ Cesium Frequency Standard is a compact, self-contained module that produces accurate, stable and spectrally pure sinusoidal signals. The 4415 satisfies system applications for precision time and frequency in ground based satellite communication systems, telecommunication synchronization systems and satellite-based timing systems. The volume is $16 \times 20 \times 42$ cm$^3$, the mass is 15 kg and the power consumption is 50 W during warm up and 30 W during operation.

During the time frame of Cosmic Vision, other space-qualified commercial clocks, especially space-qualified optical clocks, of desired stability are likely to be developed.

## 4.5 Laser pulse timing system

A T2L2 (Time Transfer by Laser Link) timing device is on board the JASON2 satellite launched on June 20, 2008. Here we adopt the specifications of this instrument. This timing instrument will be adapted to record the time position of each laser pulse (event). The precision of the timing is 2 ps.

The whole timing instrument (a timing box, $280 \times 230 \times 180$ mm, weight: 8 kg, power: 48 W) is composed of:

- Single Photon Detector (see Figure 7 and Figure 8)
- Event Timer
- Control Electronics
- Power Supply

The timing detection is asynchronous, i.e. no transponding is needed for the incoming light and outgoing light. The maximun aquisition rate is 5 kHz. The receiving times and transmitting times are recorded by the timing system and compared with those recorded in the ground station to sort out the corresponding transmitting time and receiving time of a single pulse. The onboard clock and ground clock synchronization procedure is as T2L2 using these two one-way ranging time. The uncertainty of single photon timing due to width of the pulse (50 ps) is 25 ps. With enough photons (313-625), the uncertainty can reach 1-2 ps. Including clock uncertainty, we can safely assume timing uncertainty for a normal point to be 3 ps.

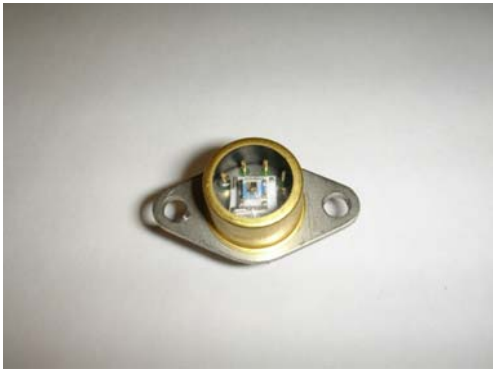 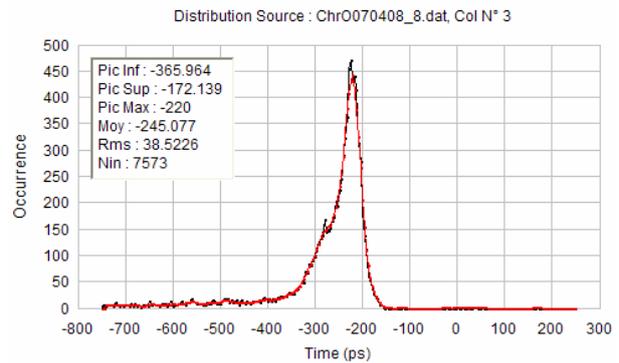

**Fig. 7** Single photon detector, quantum efficiency $\rho = 0.4$ @ 532 nm, diameter = 40 μm

**Fig. 8** Chronometry distribution in a single photon mode (ordinate: number of occurrences)

### 4.6 Inertial sensor - drag free proof mass

The spacecraft is stabilized in three axes and has to follow a geodetic orbit. For this purpose a high-precision inertial sensor is needed to detect non gravitational forces. It is part of the closed loop drag free system that forces the spacecraft to follow the geodetic orbit, which is essential to this mission. The output of the inertial sensor is used as the input for the drag free closed loop that controls the μN-Thrusters (FEEPs). The drag-free requirement is $1 \times 10^{-14}$ m s$^{-2}$ Hz$^{-1/2}$ at 300 μHz with 1/f increase at lower frequencies and f$^2$ increase above 3 mHz for the two axes of the orbital plane and less by a factor of one order of magnitude for the third axis. This performance is 10 times less stringent than the LISA drag-free system requirement at 100 μHz. The inertial sensor therefore has to fulfill as a minimum this requirement.

Because the specifications are relaxed along one axis, the inertial sensor can be optimised for the two other axes. The proof mass of the sensor ($m_p = 1.75$ kg) is a rectangular parallelepiped ($50 \times 50 \times 35$ mm$^3$) made from Au-Pt alloy of extremely low magnetic susceptibility ($5 \times 10^{-5}$). The six sides of the proof mass are surrounded by gold coated Ultra Low Expansion plates supporting electrodes mounted on the housing for capacitive sensing of the six degrees of freedom of the mass. The same electrodes are also used for the mass control according to the spacecraft operation mode and drag-free system requirements. The charge management control of the mass takes also advantage of these electrodes sets. The gap between each side of the proof mass and the opposing electrode is fixed to 2 mm, compared to the 600 μm of the MICROSCOPE [51] mission instrument and the LISA sensor with 4 mm distance. This value is a compromise between the performance of the sensor and its operation robustness. The tight Titanium housing of the instrument core will sustain the vacuum pressure less than 10 μPa thanks to the getter device and the selection of the materials for the limitation of the outgassing after baking.

Extensive calculations have been carried out concerning the physical properties of the inertial sensor for ASTROD I, like effects of acceleration disturbances due to magnetostatic and Lorentz forces, cosmic ray and residual gas impacts, gravity gradient effects, radiometric effects, thermal radiation pressure and proof mass sensor back action disturbances (section 3.4).

The technique of inertial sensing is well understood. The inertial sensors for the LISA Pathfinder mission are under development at the University of Trento in Italy (S. Vitale). Furthermore, ONERA-DMPH has flown 4 electrostatic inertial sensors during the last decade, demonstrating the interest of the concept, the possibilities of the design and the reliability of the technologies used. The GOCE satellite will be launched at the beginning of 2008 with 6 sensors of $2 \times 10^{-12}$ m s$^{-2}$ Hz$^{-1/2}$ measurement resolution used in the gravity gradiometer, in presence of more than $10^{-6}$ m s$^{-2}$ acceleration (Fig. 9): that is much more demanding than needed for the ASTROD I drag free sensor. In addition, the instrument for the MICROSCOPE mission is being produced for a test of the Equivalence Principle at $10^{-15}$ accuracy, corresponding to the in orbit detection of a signal lower than $8 \times 10^{-15}$ m s$^{-2}$. Caging is necessary for the proof mass during launch. This should be studied with the experience of LISA Pathfinder.

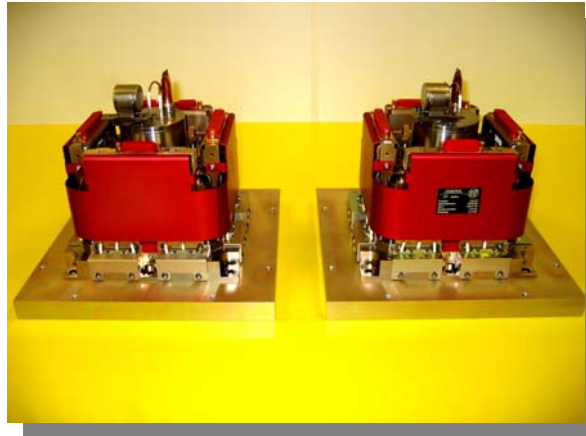

**Fig. 9** Two GOCE sensor heads (flight models) of the ultra-sensitive accelerometers (ONERA's courtesy)

### 4.7 Thermal control and diagnostic

Studies on thermal requirements and diagnostic have already been done for LISA and LISA Pathfinder. LISA Pathfinder requires a thermal stability of the order of $10^{-4}$ K Hz$^{-1/2}$ in the frequency bandwidth 1 mHz $\leq f \leq$ 30 mHz for its proof masses. To monitor the thermal environment inside the spacecraft and science module, negative temperature coefficient resistance sensors (NTC) have been tested on ground. In Lobo et al. [52] a NTC sensor with temperature sensitivity of $3 \times 10^{-5}$ K Hz$^{-1/2}$ down to frequencies near and above 0.1 mHz is reported. The current objective for the LTP diagnostic subsystem is to employ thermal sensors with temperature sensitivities of the order of $10^{-5}$ K Hz$^{-1/2}$ down to 1 mHz. For ASTROD I and taking in perspective the future implementation of the ASTROD mission concept, these requirements should be improved and extended at low frequencies, down to 0.1 mHz for ASTROD I and down to micro hertz level for the ASTROD mission. Research in that direction needs to be performed in the assessment study.

### 4.8 μN - thrusters

The μN-Thruster technology is a key technology for ultra precise drag free control systems needed to achieve the required drag free quality. The thrusters are used to compensate all non gravitational force disturbances acting on the spacecraft like thermal radiation pressure, solar pressure and other non gravitational interactions of the spacecraft with the environment. The technology of Field Emission Electrical Propulsion (FEEP) is a promising method to generate thrust forces in the μN- or even sub-μN-range (Fig. 10). For this purpose ions are accelerated by high voltages (8 kV) to velocities of several dozens of km/s. The specific impulse (ISP) of such a device is very high, i.e. 4000 to 8000 s, and so this technology is very efficient for deep space missions. The corresponding reaction of the spacecraft depends on its mass and inertia. For the MICROSCOPE mission, the resolution requirement for the FEEPs is 0.1 μN with a full range of 150 μN per thruster; the twelve thrusters are associated in four clusters of three thrusters. For ASTROD I, the configuration is rather similar with 12 thrusters too (Fig. 11); because of the more limited disturbing surface forces applied on the spacecraft, the requirement for the full range can be relaxed by one order and the resolution of the FEEP is not a major challenge because of the high gain of the drag-free servo-loop.

The main constituents of the FEEP subsystem are:

- **Thruster Assembly (TA), consisting of**
  -- the FEEP thrusters (either slit of several mm length with Cesium propellant or needle with Indium)
  -- the high energy ion source which produces the desired thrust, with its propellant tank
  -- its thermal insulation
- **Neutralizer Assembly (NA)**

The low energy electron source that balances the charge accumulated by the ion emission

- **Power Control Unit (PCU)**

The electronic box that provides power for all electronics, the high voltage sources and the operation controls of the thruster and the neutralizer.

There are 3 clusters of 4 TA on the spacecraft, each with its own PCU and NA. A single PCU drives 4 TA, two NA neutralize the charge of 4 TA.

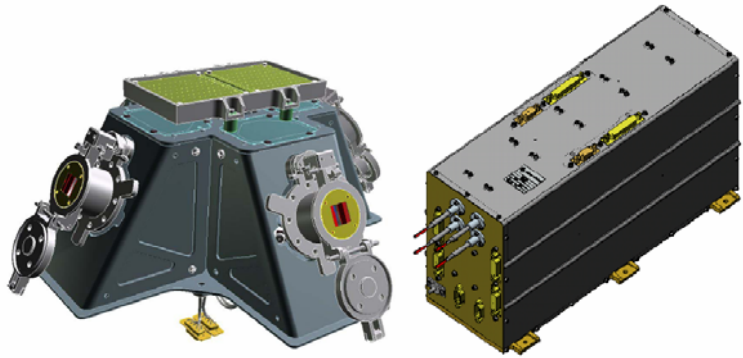

**Fig. 10** Configuration of the cluster of 4 TA with its PCU (with ESA courtesy)

The FEEP technology is currently under development at Alta/Italy for MICROSCOPE and LISA-Pathfinder. The main critical issues of the FEEP technology are the reliability of the switch on operation and the emitter's endurance which is still to be demonstrated at the level of the mission requirements, mainly linked to strict selection procedures to be tested and finalized by the life-test, clustering effects, failure modes (avalanche effect) and the lack of long term operation qualification. For MICROSCOPE cold gas thrusters are being considered as an alternative. As the requirements for ASTROD I for drag free control are comparable with that of MICROSCOPE and LISA-Pathfinder, the finally chosen technology for these missions could also be used for ASTROD I.

Independently of the technology used, the μN-thrusters are needed to control the spacecraft in all degrees of freedom, orbit motion and attitude. The plume of the thrusters, located at the outside of the spacecraft will not interact with it because of the relatively simple shape of ASTROD I.

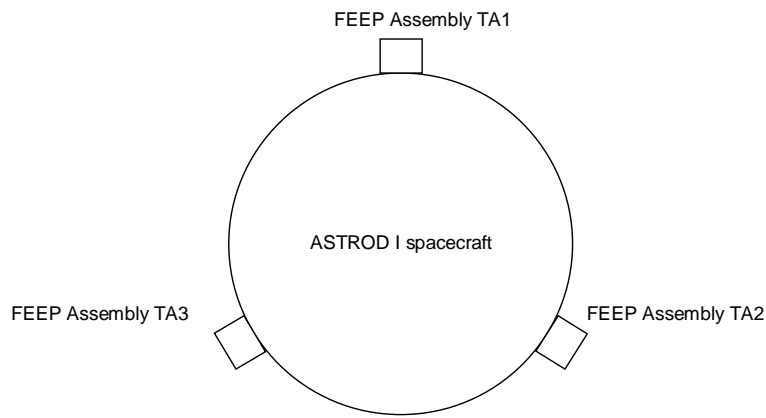

**Fig. 11** FEEP configuration. Each Assembly contains 4 FEEP thrusters

### 4.9 Charge management system

In order to avoid the charge on the proof mass exceeding the budget allowed, a discharging system is needed that discharges the proof mass using UV-radiation. The fundamental idea for charge management for ASTROD I is based on using UV light to discharge the proof mass via photoelectric emission. Such a system has been demonstrated in flight by the GP-B (Gravity Probe B) mission [53].

A photoelectric discharge system has also been developed by Imperial College London for LISA Pathfinder using UV lamps. This system has been extensively tested, both via simulations and laboratory tests [54]. A similar system could easily be used for ASTROD I. The proof mass will nominally be continuously discharged. This offers the advantage of lowering the mean discharging rate, and reducing the magnitude of coherent signals that appear due to the accumulation of proof mass charge. However, changes in the mean charging rate will limit the level to which the charging and discharging rates can be matched and hence how well these signals can be removed [55]. Relevant fluctuations in the mean charging rate are currently being investigated [56]. Although UV lamp charge management system is the baseline for ASTROD I, UV LED charge management system may also be developed for this purpose for lighter weight and larger bandwidth.

### 5. Basic Spacecraft Key Factors

As it is stated in the scientific objectives in section 2, the mission is performed by one S/C and the ground segment. For full science operation, only one ground station at one time is needed, though the particular physical station may change with Earth rotation.

The payload of the S/C collects, processes and transmits the laser data for performing the science operations. In this section the S/C bus, which provides the necessary services to enable the payload to operate, will be described. Table 5 compiles the mass and power budget of the spacecraft.

Table 5 Spacecraft mass and power budget

| Subsystem | Mass [kg] | Operational Power [W] |
|---|---|---|
| Payload | 160 | 280 |
| AOCS | 70 | 68 |
| Telecommunication | 20 | 20 |
| OBDH | 15 | 22 |
| Structure | 35 | - |
| Power | 80 | - |
| Cabling | 10 | - |
| Apogee motor | 100 | - |
| Total | 490 | 390 |

### 5.1 Attitude and orbit control system (AOCS)

As it can be seen in the mission profile in section 3, the S/C is put into a transfer orbit of large eccentricity with a perigee of about 200-300 km and an apogee of about 500,000 km by the launcher and afterwards in the first Venus approach orbit by a 100 kg liquid propulsion unit. Any attitude and orbit change thereafter has to be done by the S/C itself. Due to the scientific objectives, the S/C will be three-axis stabilized. There are three major classes of requirements, the AOCS has to fulfil.

The first class contains the injection correction and all other orbit correction manoeuvres after the propulsion unit is separated. The second class comprises any attitude control manoeuvres, that can't be conducted by the drag free control system. The third class stands for performing the drag free control.

For the first class a medium sized ion thruster is foreseen. For the first orbit insertion correction, a $\Delta v$ in the order of 0.1-1 m/s is expected. Taking into account possible further orbit correction manoeuvres, like the ones prior and after the gravity assists (although not designed), a medium sized ion thruster (40 kg) of total capability of $\Delta v = 100$ m/s will be used. Small ion thrusters could also be considered.

A number of 12 cold gas thrusters is foreseen for the second class, providing a proven and redundant design. It is needed for all attitude corrections which are beyond the capabilities of the drag free control system. Typical events will be the establishment of Sun acquisition after the separation from the launcher or the provision of the right pointing for the ion engine operation.

The thrusters for the drag free control are considered to be part of the payload and are described in section 4.8.

## 5.2 Telemetry, science data and OBDH (On-Board Data Handling)

During the mission lifetime the demand for telemetry and science data communication capabilities is moderate and explained in this section.

After the orbit injection correction, the science mission starts with a proposed life time of about 3 years. The S/C is in drag-free phase, containing two Venus (drag-free) flyby period. Before the science phase starts, in the orbit correction phase (a couple of months), the SC occasionally needs to transmit telemetry data and to receive commands, e.g. for correction manoeuvres or software updates. In the science mission phase, the transmission of science data is alsorequired. The demand for communication amounts to an approximate data rate of 5 kbit/s (assuming RF ground coverage of about 9 hours in two days).

The range between the S/C and Earth during science phase will be at maximum 1.7 AU. Thus a 1.3 m Cassegrain high gain antenna is foreseen for nominal X-band communication with the ground station. There are two additional low gain antennas for providing all-time S-Band communication capabilities. The LGAs (Low Gain Antennas) will be used for contact during the first days of the mission and any low data rate occasions during the non-science phase.

The requirements are low for on board data handling. The above mentioned demand originates from an internal data rate of about 1000 bit/s. With a regular downlink of six hours in two days, the requirement for data recording on the S/C is in the order of 170 Mbits. The processing and management of the S/C functions, like the drag free control, attitude determination or the management of the internal and external data flow is handled by the on board computing system. The requirements for ASTROD I are within the capabilities of the RAD 6000-SC (Radiation Hardened IBM RISC 6000 Single Chip) standard system which will be used for the payload and for the S/C central processor each.

## 5.3 Ground segment

The ground segment operations are divided into two topics. These are the laser ranging and the RF communication, which include the science data, telemetry and command.

As it is stated in the science objectives in section 2, the laser operations are planned for several times a day with duration in the order of 1000 s each.

This requires the utilisation of at least two laser ground stations. Two of them will be represented by the Yunnan observatory in China and the Côte d'Azur Observation station. The relevant ground segment requirements can be found in section 3.3.

The radio frequency communications have a different ground coverage demand. The basic task is to download all the telemetry and science data, which is recorded on the S/C. This can be done at any time, as long as the on board memory has enough capacity left for storing new data. The requirements on the ground coverage are approximately three hours a day with downlink data rate 8 kbit/s in a distance of 1.7 AU. This can be achieved with a single telescope in the 30 m class, being available at one of the two ESA sites DSA-1 (Australia) and DSA-2 (Spain) or the Deep Space Network.

## 6. Science Operations and Archiving

### 6.1 Science operations

Science operations will be conducted by the Science Operations Team (SOT), under the responsibility of the ESA Project Scientist, with strong support of the PI teams. The SOT will be located at a dedicated Scientific Operation Centre (SOC) during critical phases of the mission. Co-location with the MOC (Mission Operation Centre) of ESOC is recommended. For routine operations (e.g. interplanetary cruise) the SOT will interface with the MOC from its home institution.

The European Space Operations Centre (ESOC) will perform orbit determination by European station, for the purpose of rough spacecraft navigation. The radiometric tracking data, optical ranging and data from the onboard accelerometer will be collected and calibrated at the active data site.

The science operations include precise orbit determination using independent data sources in order to test the gravity law at a range of distances and during Venus flybys. This requires post-processing of the accelerometer data to calibrate instrument biases and to determine non-gravitational forces acting on the spacecraft. Then, thanks to the unbiased accelerometer data and tracking data an accurate orbit determination will be performed by the SOT, eventually with redundancy between the different institutes. The comparison of the radiometric, the laser tracking and accelerometer observables with the acceleration determined from the standard gravity model will determine if there are any discrepancies, and limit the number and parameters of possible gravity models. Orbital solutions, together with documentation of the underlying models, will be available from the active data site. This serves a twofold purpose:

1) Ensure that the orbital determination software of each science team provides consistent results.
2) Provide a solution for team-members who do not wish to perform orbital determination themselves.

These solutions will be provided and documented by the Science Operation Team.

### 6.2 Archiving

According to the ESA policy on data rights, for the first six months after the end of the mission, the team of ASTROD investigators will have exclusive rights over their data. Thereafter, all science data (raw, calibrated, housekeeping) will have to be submitted to the ASTROD Science Data Archiving Centre (ASDAC) where the data will be stored on CD-ROM and can be accessed by the wide scientific community. The complete ASTROD data set comprises:

- laser ranging data
- inertial sensor data
- drag-free control and µN- thruster data
- UV control lamp discharging data

The teams providing the various data sets have the following tasks:

- performing a thorough end-to-end error analysis
- calibration of the science data
- development of appropriate software for data analysis
- production of an explanatory supplement
- timely (i.e. 6 months after mission end) delivery of the items above to the ASDAC

The ASDAC has the following tasks:

- ensuring timely delivery of the items above
- verification of the contents of the CD-ROMs
- production of an appropriate number of copies of CD-ROMs and supplements
- responding to requests from the user community and sending out CD-ROMs and supplements as requested.

## 7. Outlook

The field of the experimental gravitational physics stands to be revolutionized by the advancements in several critical technologies, over the next few years. These technologies include deep space drag-free navigation and interplanetary laser ranging. A combination of these serves as a technology base for ASTROD I. ASTROD I is a solar-system gravity survey mission to test relativistic gravity with an improvement in sensitivity of over 3 orders of magnitude, improving our understanding of gravity and aiding the development of a new quantum gravity theory; to measure key solar system parameters with increased accuracy, advancing solar physics and our knowledge of the solar system; and to measure the time rate of change of the gravitational constant with an order of magnitude improvement and the anomalous Pioneer acceleration, thereby probing dark matter and dark energy gravitationally. This will be the beginning of a series of precise space experiments on the dynamics of gravity. The techniques are becoming mature for such experiments.

We are grateful to the referees for helpful suggestions to make our presentation clearer.